\documentclass[twocolumn,showpacs,prlprintnumbers,amsmath,amssymb]{revtex4}
\usepackage{amsmath}
\usepackage{graphicx}
\usepackage{dcolumn}
\usepackage{bm}
\usepackage{overpic}
\usepackage{booktabs}
\usepackage{color}
\usepackage{placeins} 

\def\degree{\hbox{$^\circ$}}
\newcommand{\enn}{El-Ni\~{n}o}
\newcommand{\en}{El-Ni\~{n}o }

\begin{document}

\title{Oceanic El-Ni\~{n}o wave dynamics and climate networks}
\author{Yang Wang$^1$\footnote{wangyang.maple@gmail.com}, Avi Gozolchiani$^2$, Yosef Ashkenazy$^2$, and Shlomo Havlin$^1$\footnote{havlin@ophir.ph.biu.ac.il}}

\affiliation{1 Department of Physics, Bar-Ilan University, Ramat-Gan 52900, Israel\\
  2 Solar Energy and Environmental Physics, Blaustein Institutes for Desert Research, Ben-Gurion University of the Negev, Israel
 }

\date{\today}

\begin{abstract}

Oceanic Kelvin and Rossby waves play an important role in tropical climate and \en dynamics. Here we develop and apply a climate network approach to quantify the characteristics of \en related oceanic waves, based on sea surface height satellite data. We associate the majority of dominant long distance ($\geq 500$ km) links of the network with (i) equatorial Kelvin waves, (ii) off-equatorial Rossby waves, and (iii) tropical instability waves. Notably, we find that the location of the hubs of out-going ($\sim 180^{\circ}\rm{E}$) and in-coming ($\sim 140^{\circ}\rm{W}$) links of the climate network coincide with the locations of the Kelvin wave initiation and dissipation, respectively. We also find that this dissipation at $\sim 140^{\circ}\rm{W}$ is much weaker during \en times. Moreover, the hubs of the off-equatorial network coincide with the locations of westerly wind burst activity and high wind vorticity, two mechanisms that were associated with Rossby waves activity. The quantitative methodology and measures developed here can improve the understanding of \en dynamics and possibly its prediction.

\end{abstract}
\maketitle


The \enn-Southern Oscillation (ENSO) is the largest climatic cycle on annual time scales, and is one of the most important processes that affect the natural climate variability.  \en has a profound effect not only on the tropical atmospheric and oceanic conditions, but also on remote regions including North America, Indian monsoon region, and Antarctica~\cite{Turner2004,Clarke2008,Sarachik2010}. Classical theories of \en attribute its self-sustained dynamics mainly to positive and negative feedbacks between the depth of the thermocline, sea-surface temperature, and winds.

Oceanic waves play an important role in \en dynamics, and the equatorial region can be regarded as a natural waveguide of several degree latitude width centered at the equator~\cite{Gill1982}. Tropical waves, such as Kelvin and Rossby waves~\cite{Gill1982}, are propagating in this waveguide. The intraseasonal Kelvin waves have been extensively studied using various observational and modeling methods due of their importance to the theory of El-Ni\~{n}o~\cite{Roundy2006}. For example, the 1997/8 strong \en event was initiated by a group of very pronounced Kelvin waves which were generated by energetic westerly wind bursts (WWBs)~\cite{McPhaden1999}. Recently, satellite altimetry data has become an invaluable resource to study equatorial wave dynamics~\cite{Chelton2003, Wakata2007}. The velocity of waves in the equatorial Pacific region can be estimated using the so-called Hovm\"oller (longitude-time) diagrams, by transferring the data to the wavenumber-frequency space, or by decomposing the data into several leading modes of equatorial Kelvin and Rossby waves~\cite{Delcroix1994,Wakata2007}. Generally speaking, the typical phase speeds of the first baroclinic mode of equatorial Kelvin and Rossby waves are $\sim 2.8 \rm{m/s}$ and $\sim 0.9 \rm{m/s}$, respectively~\cite{Delcroix1994,Wakata2007} where Kelvin waves exist on the equator and propagate eastward while Rossby waves are off-equatorial waves that propagate to the west. 

Tropical oceanic waves connect different, sometimes very far, regions. The set of regions $V$ and their interactions $E$ can be encoded in a graph $G(V,E)$ named in recent literature ``climate network''~\cite{Tsonis2008}. Many fields of research benefit from a rich set of quantifications and algorithms that has been developed for analyzing such networks of interactions. Examples include social and biological systems, information flow through the world wide web and physiological activities (see~\cite{Newman2010, Cohen2010} for reviews). In climate science, network theory has been used, e.g., to infer known climate patterns (like ocean currents), to investigate atmospheric variabilities, and to study low-frequency climate phenomena~\cite{Tsonis2008,Yamasaki2008,Donges2009,Wang2013,Gozolchiani2011,Berezin2012,Ludescher2013,Feng2014a}. It is common to regard links from $E$ whose statistical significance is above a certain threshold as weighted or unweighted links. Since network based methods emphasize coordination between locations rather than local dynamics, they exhibit an increased sensitivity to climate patterns which are not easily captured by conventional analysis~\cite{Ludescher2013,Feng2014a}. In spite of the importance of oceanic dynamics on the climate system, only a few studies have used oceanic data within the framework climate networks~\cite{Feng2014a,GRL:GRL52004}. Also, the studies of \en using climate network methodology took advantage of near surface atmospheric data~\cite{Yamasaki2008,Gozolchiani2011,Ludescher2013}. Here we aim to fill this gap, focusing on the important ocean-atmosphere phenomenon, \enn. 

By transforming the daily and relatively long ($\sim$20 years) satellite based altimeter (i.e., sea surface height) records into climate networks, we find that the time delays, the velocities and the directions associated with climate network links can be attributed to equatorial Kelvin, Rossby and tropical instability waves (TIW). As will be shown below, the topology of the climate network is strongly affected by \enn, which implies that the relative number of links in the eastern equatorial Pacific is significantly higher during \en times compared to normal times. Notably, the network properties reveal the locations of the wave initiation as well as its strong dissipation. Our results suggest that climate network may serve as a useful tool to study wave dynamics, and more importantly, network features may serve as their quantitative measurements.


We analyze the daily sea surface height (SSH) anomaly data~\cite{AVISO2013}. We focus on the tropical Pacific ocean between 7\degree{S} to 7\degree{N}, and between 120\degree{E} to 280\degree{E} (80\degree{W}); the time extent of the data is from January 1 1993 to December 31 2010. The spatial resolution is 1\degree{} and 4\degree{} in the meridional and zonal directions, respectively. Totally, there are 615 grid points, which are regarded as nodes of the climate network.  For each node of the network, we analyze daily SSH values after subtracting the seasonal cycle. The filtered SSH record is denoted by $H$. We also define the normalized series ${\Theta _s}(t) \equiv {{{\left[ {{H_s}(t) - \left\langle {{H_s}(t)} \right\rangle } \right]} \mathord{\left/ {\vphantom {{\left[ {{H_s}(t) - \left\langle {{H_s}(t)} \right\rangle } \right]} {\left\langle {{{\left( {{H_s}(t) - \left\langle {{H_s}(t)} \right\rangle } \right)}^2}} \right\rangle }}} \right.  \kern-\nulldelimiterspace} {\left\langle {{{\left( {{H_s}(t) - \left\langle {{H_s}(t)} \right\rangle } \right)}^2}} \right\rangle }}^{{1 \mathord{\left/ {\vphantom {1 2}} \right.  \kern-\nulldelimiterspace} 2}}}$, where $\langle \cdots \rangle$ represents the temporal average, $s$ is the location, and $t$ is the time parameter.

A link between each pair of sites $s_1$ and $s_2$ on year $y$ is defined based on the lagged cross-correlation function $X_{{s_1},{s_2}}^y(\tau \ge 0) = \left\langle {\Theta _{{s_1}}^y(t)}{\Theta _{{s_2}}^y(t + \tau )} \right\rangle $, and $X_{{s_1},{s_2}}^y(\tau)=X_{{s_2},{s_1}}^y(-\tau)$, where $\tau$ is the time lag parameter that is considered up to 200 days (the results are not sensitive to this parameter). We also define the optimal time lag, $\tau^*$, at which $X_{{s_1},{s_2}}^y(\tau)$ is maximal (or minimal), as the time delay of a pair $s_1$, $s_2$. When $s_1$ is to the west of $s_2$ and the time lag is positive, the link direction is to the east. We distinguish between positive and negative link {\it weights} as follows
\begin{equation}
W_{{s_1},{s_2}}^{y,+} = \frac{{{\rm max}(X_{{s_1},{s_2}}^y) - {\rm
      mean}(X_{{s_1},{s_2}}^y)}}{{{\rm std}(X_{{s_1},{s_2}}^y)}},
\label{equ1}
\end{equation}
and,
\begin{equation}
W_{{s_1},{s_2}}^{y,-} = \frac{{{\rm min}(X_{{s_1},{s_2}}^y) - {\rm
      mean}(X_{{s_1},{s_2}}^y)}}{{{\rm std}(X_{{s_1},{s_2}}^y)}},
\label{equ2}
\end{equation}
where ${\rm max}$ and ${\rm min}$ are the maximum and minimum values of the cross-correlation $X_{{s_1},{s_2}}^y(\tau)$, ${\rm mean}$ and ${\rm std}$ represents the mean and standard deviation. Each climate network is constructed based on 1~yr SSH data, and the time resolution for two consecutive climate network is 30 days. To eliminate the effect of slow modulations on correlation estimates, a two years high-pass filter has been used. Examples of SSH time series and their cross correlation-function is presented in Fig. S1 of the SI. 

\begin{figure*}[htbp]
\centering
\includegraphics[width=0.8\textwidth]{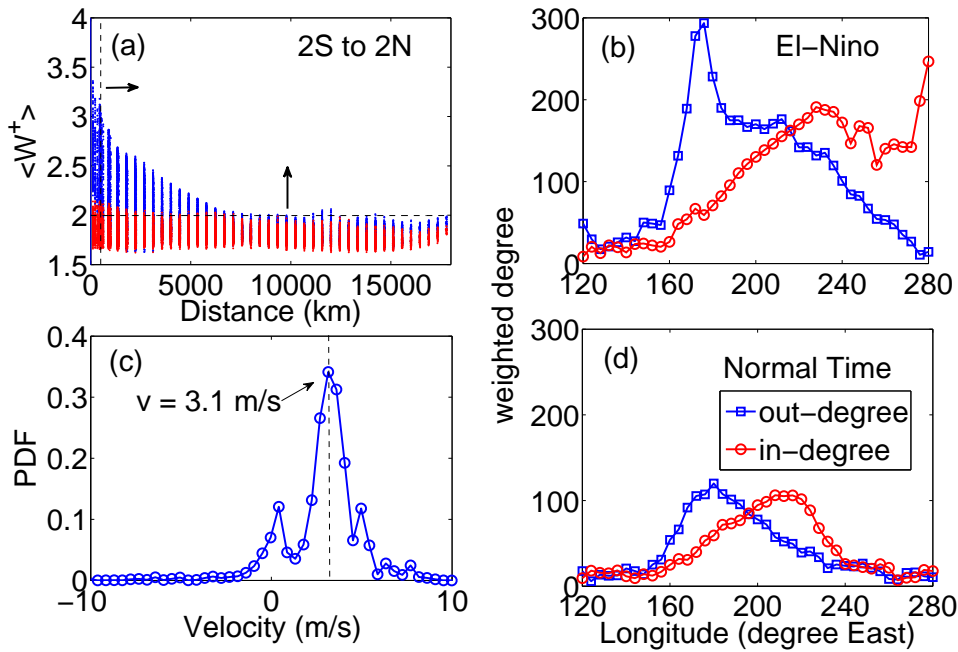}
\caption{
  Equatorial Pacific network (2\degree{N} to 2\degree{S}) analysis. (a) The temporal mean weight of positive links $\langle W^{+} \rangle$ versus their distance for both real (blue dots) and shuffled (red dots) data. (b) The ``in'' (red circles) and ``out'' (blue squares) weighted degree as a function of longitude for \en times, for positive links ($p \le 0.02$ and $D \geq 500 \rm{km}$). (c) The probability density function (PDF) of phase speeds that are based on positive links with $p \le 0.02$ and distances larger than $500 \rm{km}$. (d) Same as (b) but for normal times.}
 \label{Fig1}
\end{figure*}

We construct climate networks for three sub-regions of the tropical Pacific: equatorial Pacific (2\degree{S} to 2\degree{N}), northern tropical Pacific (4\degree{N} to 7\degree{N}) and southern tropical Pacific (4\degree{S} to 7\degree{S}). To identify a threshold below which link weights $W$ can be disregarded, we apply a shuffling procedure in which the order of years is shuffled while the order of days within each year remains unchanged. Thus, statistical properties of the data such as the distribution of values and their autocorrelation functions (within 1 year) are not affected by the shuffling procedure, but the statistical dependence between different nodes is diminished. Fig.~\ref{Fig1}a depicts the temporal average positive link weights, $\langle W^{+} \rangle$, versus the geographical distances of the links $D$, for the real and shuffled data in the equatorial Pacific regions (from 2\degree{S} to 2\degree{N}). High positive average link weight values which exist in the real data but not in the shuffled data are less likely to occur by chance. We thus consider only links that are separated from the shuffled links (indicated by the horizontal dashed lines and vertical arrows in Fig.~\ref{Fig1}a); in addition, short distanced links (indicated by the vertical dashed line and the horizontal arrow in Fig. 1a) exhibit trivial correlations and are thus ignored (the ``proximity effect''~\cite{Wang2013}). The probability density function (PDF) of average link weight in this region can be seen in Fig. S2 of the SI.  We find that significant links with a {\it p}-value $p\le0.02$ include $26.6\%$ of the weighted links in the equatorial Pacific (2\degree{S} to 2\degree{N}), $6.5\%$ for 4\degree{N} to 7\degree{N}, and $11.7\%$ for 4\degree{S} to 7\degree{S}.

One of the main characteristics of equatorial Kelvin waves is their phase speed (which is equal to their group velocity since these waves are non-dispersive)--below we use the links' statistics to estimate this velocity. Assuming that the delay time $\tau^*$ is a good estimator for a dynamical interaction time between two nodes~\cite{Cimponeriu2004}, the phase velocities can be estimated by dividing the geographical distance $D$ between the nodes by the delay time $\tau^*$, i.e., $v \approx D/\tau^*$. The PDF of phase speeds (of links with $D\geq 500$ km and $p\le 0.02$) is shown in Fig.~\ref{Fig1}c. For the equatorial Pacific, the phase speed is positive (eastward) and peaks around $3 \rm{m/s}$, in agreement with the value and direction of Kelvin waves~\cite{Toshiaki2008,Wakata2007}. 

The topological properties of the climate network in equatorial Pacific region show distinct behavior during \en and normal times. The climate network considered above is directed---positive and negative $\tau^*$ indicates eastward and westward flow respectively. We distinguish between a link that is pointing towards a node, or away from the node; the former/ latter is related to the total link weight pointing to/ from a specific node and is referred to below as weighted ``in''/ ``out''-degree.  Fig.~\ref{Fig1}b,d depict the ``in'' and ``out'' weighted degree versus longitude, averaging over all the latitudes from 2\degree{S} to 2\degree{N} during \en and normal times, respectively.  First, we find that the ``out''-weighted degree is peaked around 170\degree{E} at the equator with a much more pronounced peak during \enn, which indicates enhanced Kelvin waves activity, consistent with previous, more conventional, analysis~\cite{McPhaden1999}. Indeed, stronger Kelvin wave activity during \en around 170\degree{E} was associated with strong WWBs~\cite{Eisenman2005}, which is enhanced wind activity that underlies the generation of equatorial Kelvin waves. Second, there is a peak in the ``in''-weighted degree curve at $\sim$220\degree{E} during both \en and normal times, where for \en times the ``in'' degree curve is high all the way to the coast of South America while during normal times the ``in''-degree curve decays for longitudes larger than $\sim$220\degree{E}. We interpret this decay as strong dissipation of the Kelvin waves, and our results suggest that the dissipation of Kelvin waves is much weaker during \enn. 

The attenuation of the Kelvin waves to the east of $\sim$220\degree{E} during \en was implied based on wave packets (found using Hovm\"oller diagrams) using observation and model data~\cite{QJ:QJ200212858211,JGRC:JGRC20892}. This attenuation may be attributed to several mechanisms, including the strong vertical current shear during La-Ni\~{na}, strong wind at $\sim$220\degree{E}~\cite{QJ:QJ200212858211}, scattering of waves, and partial reflection of Kelvin waves into off-equatorial Rossby waves~\cite{JGRC:JGRC20892}.  Here, we use the climate network approach to quantify the waves properties in a statistical and consistent way. 

\begin{figure}
\center\includegraphics[width=\columnwidth,trim=50 50 50 20,clip=false]{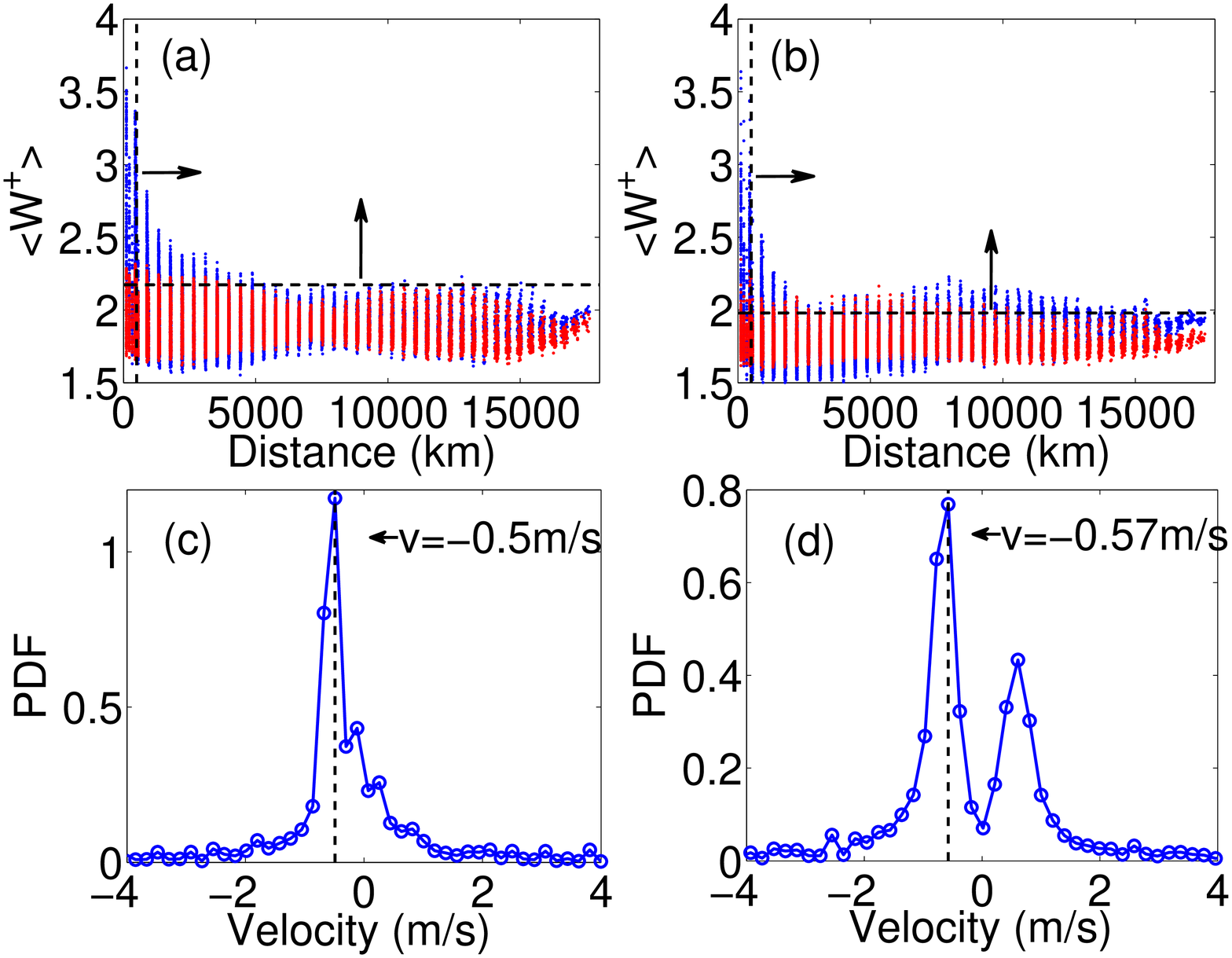}
\caption{  The temporal average link weight $\langle W^{+} \rangle$ versus geographical distance $D$ in ({a}) the northern tropical Pacific (4\degree{N} to 7\degree{N}) and ({b}) the southern tropical Pacific (7\degree{S} to 4\degree{S}). The PDF of phase velocity in ({c}) the northern tropical Pacific and ({d}) the southern tropical Pacific.}
\label{Fig2}
\end{figure}

\begin{figure}
\center\includegraphics[width=\columnwidth,trim=20 10 20 20,clip=true]{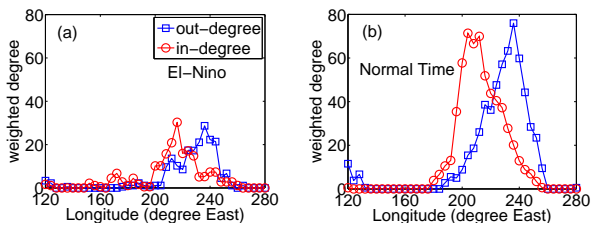}
\caption{The out- (blue squares) and in- (red circles) weighted degree of nodes in the climate network from 4\degree{N} to 7\degree{N} during ({a}) El-Ni\~{n}o times and ({b}) normal times.}
\label{Fig3}
\end{figure}

We next apply the climate network methodology to the northern and southern tropical Pacific to study Rossby (and other) tropical waves,  respectively. The statistical properties of positive links in the climate networks in the northern tropical Pacific (4\degree{N} to 7\degree{N}) and southern tropical Pacific (4\degree{S} to 7\degree{S}), are shown in Fig.~\ref{Fig2}. The phase speed that is calculated based on links above significance level ($p \le 0.02$ and $D \geq 500 \mathrm{km}$) is negative (westward) and their PDF peaks around $0.5 \mathrm{m/s}$ and $0.6 \mathrm{m/s}$ respectively (Fig.~\ref{Fig2}c,d), in general agreement with the phase velocity of tropical instability waves (TIW) and off-equatorial Rossby waves~\cite{Chelton2003,Kessler1990,Wakata2007,Chelton1996,Willett2006,Lyman2005}. The structure of link weight versus distance is different for the northern and southern tropical Pacific (Fig.~\ref{Fig2}a,b); see \cite{Chelton2003} and supporting information (Fig. S3 and Fig. S4). 
Network analysis based on northern tropical Pacific is presented in Fig.~\ref{Fig3}. The westward propagation of waves is evident during both normal and \en times, with most of the wave activity concentrated between 200\degree{E} to 250\degree{E} (see also~\cite{Chelton2003}).   Evidently, extra-tropical waves are less pronounced during \en times (Fig.~\ref{Fig3}).  The enhanced wave activity during normal times can be associated with the TIW. The characteristic phase speed, location and wave activity during \en of TIW indeed fits the network observations~\cite{Willett2006}. 

\begin{figure}

\center\includegraphics[width=0.8\columnwidth,trim=20 0 20 20,clip=true]{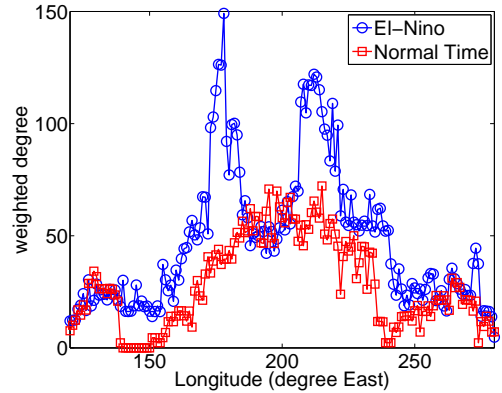}
\caption{The weighted degree in 4\degree{N} for \en (blue circles) and normal (red boxes) times. }
\label{Fig4}
\end{figure}

The TIW are not directly related to the theory of \en even though they are possible sources of random perturbations that affect both the mean state and interannual climate variations. Therefore, we filter them out by employing a sixty days low-pass filter on the SSH data as their characteristic period is about 30 days, which is much smaller than the period of Rossby waves~\cite{Lyman2005}. After this filtering, we reconstruct the climate network and calculate the corresponding wave characteristics (see Figs. S5, S6 of the SI). We relate this ``filtered'' climate network with Rossby waves, which play an essential role in the self-sustained \en dynamics. Fig.~\ref{Fig4} depicts the weighted degree at 4\degree{N} during \en and normal times. 
The Rossby waves signal, represented here by the weighted degree of the climate network, is only clearly observed mainly to the west of $\sim$ 240\degree{E} (120\degree{W}) (see Fig.~\ref{Fig4}), although some of Rossby waves close to the eastern boundary ($\sim$ 280\degree{E}) are the reflection of Kelvin waves. Note that Rossby waves are more pronounced during \en times: Around the date line (180\degree{E}) there is a peak in the weighted degree, which can be associated to the strong WWBs activity in this region. In addition, there is another peak in the weighted degree distribution, between $\sim$200\degree{E} to $\sim$230\degree{E}, which can be associated with the enhanced wind-stress curl in this region. This supports the recent suggestion that the enhanced wind-stress curl is related to the Rossby wave activity~\cite{Chelton2003}.

In summary, we construct and analyze the climate network of the tropical Pacific ocean based on daily satellite sea-surface height data (altimetry). We find that the most dominant long distance links in the network share the same characteristics (speed and direction) as equatorial Kelvin waves, off-equatorial Rossby waves, and tropical instability waves, indicating that climate network approach is useful to capture and quantify oceanic wave activity. This conclusion is further strengthened by the observation that the location of hubs in network coincide with the location of initiation and dissipation regions of equatorial oceanic waves. Moreover, the climate network shows distinct features during \en and normal times, such as the pronounced equatorial Kelvin waves, off-equatorial Rossby waves, and suppressed TIW during \enn. In addition, the dissipation of Kelvin waves are much weaker during \enn. 
This newly developed methodology based on long-term data (18 years) could serve as a quantitative tool to improve our understanding of the underlying mechanisms of climate phenomena like \enn.

Recent studies have demonstrated the ability of climate network based-techniques to predict \en events~\cite{Ludescher2013,Ludescher11022014}. This technique is based on atmospheric data above the sea surface, although as shown here the ocean plays a central role in \en phenomenon. Based on the present study we conjecture that the inclusion of oceanic data into network analysis will improve the prediction power of network techniques.

\section*{ACKNOWLEDGMENTS}
The authors would like to acknowledge the support of the LINC project (no. 289447) funded by the EC's Marie-Curie ITN program (FP7-PEOPLE-2011-ITN), Multiplex (No. 317532) EU projects, the DFG and the Israel Science Foundation for financial support. The altimeter products were produced by Ssalto/Duacs and distributed by Aviso with support from Cnes.

\begin{thebibliography}{31}%
\makeatletter
\providecommand \@ifxundefined [1]{%
 \@ifx{#1\undefined}
}%
\providecommand \@ifnum [1]{%
 \ifnum #1\expandafter \@firstoftwo
 \else \expandafter \@secondoftwo
 \fi
}%
\providecommand \@ifx [1]{%
 \ifx #1\expandafter \@firstoftwo
 \else \expandafter \@secondoftwo
 \fi
}%
\providecommand \natexlab [1]{#1}%
\providecommand \enquote  [1]{``#1''}%
\providecommand \bibnamefont  [1]{#1}%
\providecommand \bibfnamefont [1]{#1}%
\providecommand \citenamefont [1]{#1}%
\providecommand \href@noop [0]{\@secondoftwo}%
\providecommand \href [0]{\begingroup \@sanitize@url \@href}%
\providecommand \@href[1]{\@@startlink{#1}\@@href}%
\providecommand \@@href[1]{\endgroup#1\@@endlink}%
\providecommand \@sanitize@url [0]{\catcode `\\12\catcode `\$12\catcode
  `\&12\catcode `\#12\catcode `\^12\catcode `\_12\catcode `\%12\relax}%
\providecommand \@@startlink[1]{}%
\providecommand \@@endlink[0]{}%
\providecommand \url  [0]{\begingroup\@sanitize@url \@url }%
\providecommand \@url [1]{\endgroup\@href {#1}{\urlprefix }}%
\providecommand \urlprefix  [0]{URL }%
\providecommand \Eprint [0]{\href }%
\providecommand \doibase [0]{http://dx.doi.org/}%
\providecommand \selectlanguage [0]{\@gobble}%
\providecommand \bibinfo  [0]{\@secondoftwo}%
\providecommand \bibfield  [0]{\@secondoftwo}%
\providecommand \translation [1]{[#1]}%
\providecommand \BibitemOpen [0]{}%
\providecommand \bibitemStop [0]{}%
\providecommand \bibitemNoStop [0]{.\EOS\space}%
\providecommand \EOS [0]{\spacefactor3000\relax}%
\providecommand \BibitemShut  [1]{\csname bibitem#1\endcsname}%
\let\auto@bib@innerbib\@empty
\bibitem [{\citenamefont {Turner}(2004)}]{Turner2004}%
  \BibitemOpen
  \bibfield  {author} {\bibinfo {author} {\bibfnamefont {J.}~\bibnamefont
  {Turner}},\ }\href@noop {} {\bibfield  {journal} {\bibinfo  {journal} {Int.
  J. Climatol.}\ }\textbf {\bibinfo {volume} {24}},\ \bibinfo {pages} {1}
  (\bibinfo {year} {2004})}\BibitemShut {NoStop}%
\bibitem [{\citenamefont {Clarke}(2008)}]{Clarke2008}%
  \BibitemOpen
  \bibfield  {author} {\bibinfo {author} {\bibfnamefont {A.~J.}\ \bibnamefont
  {Clarke}},\ }\href@noop {} {\emph {\bibinfo {title} {An Introduction to the
  Dynamics of El Ni\~{n}o the Southern Oscillation}}}\ (\bibinfo  {publisher}
  {Elsevier Academic Press, London},\ \bibinfo {year} {2008})\BibitemShut
  {NoStop}%
\bibitem [{\citenamefont {E.~S.~Sarachik}(2010)}]{Sarachik2010}%
  \BibitemOpen
  \bibfield  {author} {\bibinfo {author} {\bibfnamefont {M.~A.~C.}\
  \bibnamefont {E.~S.~Sarachik}},\ }\href@noop {} {\emph {\bibinfo {title}
  {{The El Ni\~{n}o-Southern Oscillation Phenomenon}}}}\ (\bibinfo  {publisher}
  {Cambridge University Press, Cambridge, England},\ \bibinfo {year}
  {2010})\BibitemShut {NoStop}%
\bibitem [{\citenamefont {Gill}(1982)}]{Gill1982}%
  \BibitemOpen
  \bibfield  {author} {\bibinfo {author} {\bibfnamefont {A.~E.}\ \bibnamefont
  {Gill}},\ }\href@noop {} {\emph {\bibinfo {title} {{Atmosphere-Ocean
  Dynamics}}}}\ (\bibinfo  {publisher} {Academic Press},\ \bibinfo {year}
  {1982})\BibitemShut {NoStop}%
\bibitem [{\citenamefont {Roundy}\ and\ \citenamefont
  {Kiladis}(2006)}]{Roundy2006}%
  \BibitemOpen
  \bibfield  {author} {\bibinfo {author} {\bibfnamefont {P.~E.}\ \bibnamefont
  {Roundy}}\ and\ \bibinfo {author} {\bibfnamefont {G.~N.}\ \bibnamefont
  {Kiladis}},\ }\href@noop {} {\bibfield  {journal} {\bibinfo  {journal}
  {Journal of Climate}\ }\textbf {\bibinfo {volume} {19}},\ \bibinfo {pages}
  {5253} (\bibinfo {year} {2006})}\BibitemShut {NoStop}%
\bibitem [{\citenamefont {McPhaden}\ and\ \citenamefont
  {Yu}(1999)}]{McPhaden1999}%
  \BibitemOpen
  \bibfield  {author} {\bibinfo {author} {\bibfnamefont {M.~J.}\ \bibnamefont
  {McPhaden}}\ and\ \bibinfo {author} {\bibfnamefont {X.}~\bibnamefont {Yu}},\
  }\href@noop {} {\bibfield  {journal} {\bibinfo  {journal} {Geophys. Res.
  Lett.}\ }\textbf {\bibinfo {volume} {26}},\ \bibinfo {pages} {2961} (\bibinfo
  {year} {1999})}\BibitemShut {NoStop}%
\bibitem [{\citenamefont {Chelton}\ \emph {et~al.}(2003)\citenamefont
  {Chelton}, \citenamefont {Schlax}, \citenamefont {Lyman},\ and\ \citenamefont
  {Johnson}}]{Chelton2003}%
  \BibitemOpen
  \bibfield  {author} {\bibinfo {author} {\bibfnamefont {D.}~\bibnamefont
  {Chelton}}, \bibinfo {author} {\bibfnamefont {M.}~\bibnamefont {Schlax}},
  \bibinfo {author} {\bibfnamefont {J.}~\bibnamefont {Lyman}}, \ and\ \bibinfo
  {author} {\bibfnamefont {G.}~\bibnamefont {Johnson}},\ }\href@noop {}
  {\bibfield  {journal} {\bibinfo  {journal} {Progress in Oceanography}\
  }\textbf {\bibinfo {volume} {56}},\ \bibinfo {pages} {323} (\bibinfo {year}
  {2003})}\BibitemShut {NoStop}%
\bibitem [{\citenamefont {Wakata}(2007)}]{Wakata2007}%
  \BibitemOpen
  \bibfield  {author} {\bibinfo {author} {\bibfnamefont {Y.}~\bibnamefont
  {Wakata}},\ }\href@noop {} {\bibfield  {journal} {\bibinfo  {journal}
  {Journal of Oceanography}\ }\textbf {\bibinfo {volume} {63}},\ \bibinfo
  {pages} {483} (\bibinfo {year} {2007})}\BibitemShut {NoStop}%
\bibitem [{\citenamefont {Delcroix}\ \emph {et~al.}(1994)\citenamefont
  {Delcroix}, \citenamefont {Boulanger}, \citenamefont {Masia},\ and\
  \citenamefont {Menkes}}]{Delcroix1994}%
  \BibitemOpen
  \bibfield  {author} {\bibinfo {author} {\bibfnamefont {T.}~\bibnamefont
  {Delcroix}}, \bibinfo {author} {\bibfnamefont {J.-P.}\ \bibnamefont
  {Boulanger}}, \bibinfo {author} {\bibfnamefont {F.}~\bibnamefont {Masia}}, \
  and\ \bibinfo {author} {\bibfnamefont {C.}~\bibnamefont {Menkes}},\
  }\href@noop {} {\bibfield  {journal} {\bibinfo  {journal} {J. Geophys. Res.}\
  }\textbf {\bibinfo {volume} {99}},\ \bibinfo {pages} {25093} (\bibinfo {year}
  {1994})}\BibitemShut {NoStop}%
\bibitem [{\citenamefont {Tsonis}\ and\ \citenamefont
  {Swanson}(2008)}]{Tsonis2008}%
  \BibitemOpen
  \bibfield  {author} {\bibinfo {author} {\bibfnamefont {A.~A.}\ \bibnamefont
  {Tsonis}}\ and\ \bibinfo {author} {\bibfnamefont {K.~L.}\ \bibnamefont
  {Swanson}},\ }\href@noop {} {\bibfield  {journal} {\bibinfo  {journal} {Phys.
  Rev. Lett.}\ }\textbf {\bibinfo {volume} {100}},\ \bibinfo {pages} {228502}
  (\bibinfo {year} {2008})}\BibitemShut {NoStop}%
\bibitem [{\citenamefont {Newman}(2010)}]{Newman2010}%
  \BibitemOpen
  \bibfield  {author} {\bibinfo {author} {\bibfnamefont {M.~E.~J.}\
  \bibnamefont {Newman}},\ }\href@noop {} {\emph {\bibinfo {title} {{Networks:
  An Introduction}}}}\ (\bibinfo  {publisher} {Oxford University, New York},\
  \bibinfo {year} {2010})\BibitemShut {NoStop}%
\bibitem [{\citenamefont {Cohen}\ and\ \citenamefont
  {Havlin}(2010)}]{Cohen2010}%
  \BibitemOpen
  \bibfield  {author} {\bibinfo {author} {\bibfnamefont {R.}~\bibnamefont
  {Cohen}}\ and\ \bibinfo {author} {\bibfnamefont {S.}~\bibnamefont {Havlin}},\
  }\href@noop {} {\emph {\bibinfo {title} {{Complex Networks: Structure,
  Robustness and Function}}}}\ (\bibinfo  {publisher} {Cambridge University
  Press, Cambridge, England},\ \bibinfo {year} {2010})\BibitemShut {NoStop}%
\bibitem [{\citenamefont {Yamasaki}\ \emph {et~al.}(2008)\citenamefont
  {Yamasaki}, \citenamefont {Gozolchiani},\ and\ \citenamefont
  {Havlin}}]{Yamasaki2008}%
  \BibitemOpen
  \bibfield  {author} {\bibinfo {author} {\bibfnamefont {K.}~\bibnamefont
  {Yamasaki}}, \bibinfo {author} {\bibfnamefont {A.}~\bibnamefont
  {Gozolchiani}}, \ and\ \bibinfo {author} {\bibfnamefont {S.}~\bibnamefont
  {Havlin}},\ }\href@noop {} {\bibfield  {journal} {\bibinfo  {journal}
  {Physical Review Letters}\ }\textbf {\bibinfo {volume} {100}},\ \bibinfo
  {pages} {228501} (\bibinfo {year} {2008})}\BibitemShut {NoStop}%
\bibitem [{\citenamefont {Donges}\ \emph {et~al.}(2009)\citenamefont {Donges},
  \citenamefont {Zou}, \citenamefont {Marwan},\ and\ \citenamefont
  {Kurths}}]{Donges2009}%
  \BibitemOpen
  \bibfield  {author} {\bibinfo {author} {\bibfnamefont {J.~F.}\ \bibnamefont
  {Donges}}, \bibinfo {author} {\bibfnamefont {Y.}~\bibnamefont {Zou}},
  \bibinfo {author} {\bibfnamefont {N.}~\bibnamefont {Marwan}}, \ and\ \bibinfo
  {author} {\bibfnamefont {J.}~\bibnamefont {Kurths}},\ }\href@noop {}
  {\bibfield  {journal} {\bibinfo  {journal} {Europhysics Letters}\ }\textbf
  {\bibinfo {volume} {87}},\ \bibinfo {pages} {48007} (\bibinfo {year}
  {2009})}\BibitemShut {NoStop}%
\bibitem [{\citenamefont {Wang}\ \emph {et~al.}(2013)\citenamefont {Wang},
  \citenamefont {Gozolchiani}, \citenamefont {Ashkenazy}, \citenamefont
  {Berezin}, \citenamefont {Guez},\ and\ \citenamefont {Havlin}}]{Wang2013}%
  \BibitemOpen
  \bibfield  {author} {\bibinfo {author} {\bibfnamefont {Y.}~\bibnamefont
  {Wang}}, \bibinfo {author} {\bibfnamefont {A.}~\bibnamefont {Gozolchiani}},
  \bibinfo {author} {\bibfnamefont {Y.}~\bibnamefont {Ashkenazy}}, \bibinfo
  {author} {\bibfnamefont {Y.}~\bibnamefont {Berezin}}, \bibinfo {author}
  {\bibfnamefont {O.}~\bibnamefont {Guez}}, \ and\ \bibinfo {author}
  {\bibfnamefont {S.}~\bibnamefont {Havlin}},\ }\href@noop {} {\bibfield
  {journal} {\bibinfo  {journal} {Physical Review Letters}\ }\textbf {\bibinfo
  {volume} {111}},\ \bibinfo {pages} {138501} (\bibinfo {year}
  {2013})}\BibitemShut {NoStop}%
\bibitem [{\citenamefont {Gozolchiani}\ \emph {et~al.}(2011)\citenamefont
  {Gozolchiani}, \citenamefont {Yamasaki},\ and\ \citenamefont
  {Havlin}}]{Gozolchiani2011}%
  \BibitemOpen
  \bibfield  {author} {\bibinfo {author} {\bibfnamefont {A.}~\bibnamefont
  {Gozolchiani}}, \bibinfo {author} {\bibfnamefont {K.}~\bibnamefont
  {Yamasaki}}, \ and\ \bibinfo {author} {\bibfnamefont {S.}~\bibnamefont
  {Havlin}},\ }\href@noop {} {\bibfield  {journal} {\bibinfo  {journal}
  {Physical Review Letters}\ }\textbf {\bibinfo {volume} {107}},\ \bibinfo
  {pages} {148501} (\bibinfo {year} {2011})}\BibitemShut {NoStop}%
\bibitem [{\citenamefont {Berezin}\ \emph {et~al.}(2012)\citenamefont
  {Berezin}, \citenamefont {Gozolchiani}, \citenamefont {Guez},\ and\
  \citenamefont {Havlin}}]{Berezin2012}%
  \BibitemOpen
  \bibfield  {author} {\bibinfo {author} {\bibfnamefont {Y.}~\bibnamefont
  {Berezin}}, \bibinfo {author} {\bibfnamefont {A.}~\bibnamefont
  {Gozolchiani}}, \bibinfo {author} {\bibfnamefont {O.}~\bibnamefont {Guez}}, \
  and\ \bibinfo {author} {\bibfnamefont {S.}~\bibnamefont {Havlin}},\
  }\href@noop {} {\bibfield  {journal} {\bibinfo  {journal} {Scientific
  reports}\ }\textbf {\bibinfo {volume} {2}},\ \bibinfo {pages} {666} (\bibinfo
  {year} {2012})}\BibitemShut {NoStop}%
\bibitem [{\citenamefont {Ludescher}\ \emph {et~al.}(2013)\citenamefont
  {Ludescher}, \citenamefont {Gozolchiani}, \citenamefont {Bogachev},
  \citenamefont {Bunde}, \citenamefont {Havlin},\ and\ \citenamefont
  {Schellnhuber}}]{Ludescher2013}%
  \BibitemOpen
  \bibfield  {author} {\bibinfo {author} {\bibfnamefont {J.}~\bibnamefont
  {Ludescher}}, \bibinfo {author} {\bibfnamefont {A.}~\bibnamefont
  {Gozolchiani}}, \bibinfo {author} {\bibfnamefont {M.}~\bibnamefont
  {Bogachev}}, \bibinfo {author} {\bibfnamefont {A.}~\bibnamefont {Bunde}},
  \bibinfo {author} {\bibfnamefont {S.}~\bibnamefont {Havlin}}, \ and\ \bibinfo
  {author} {\bibfnamefont {H.}~\bibnamefont {Schellnhuber}},\ }\href@noop {}
  {\bibfield  {journal} {\bibinfo  {journal} {Proceedings of the National
  Academy of Sciences}\ }\textbf {\bibinfo {volume} {110}},\ \bibinfo {pages}
  {11742} (\bibinfo {year} {2013})}\BibitemShut {NoStop}%
\bibitem [{\citenamefont {Feng}\ and\ \citenamefont
  {Dijkstra}(2014)}]{Feng2014a}%
  \BibitemOpen
  \bibfield  {author} {\bibinfo {author} {\bibfnamefont {Q.}~\bibnamefont
  {Feng}}\ and\ \bibinfo {author} {\bibfnamefont {H.}~\bibnamefont
  {Dijkstra}},\ }\href@noop {} {\bibfield  {journal} {\bibinfo  {journal}
  {Geophysical Research Letters}\ }\textbf {\bibinfo {volume} {41}},\ \bibinfo
  {pages} {541} (\bibinfo {year} {2014})}\BibitemShut {NoStop}%
\bibitem [{\citenamefont {Feng}\ \emph {et~al.}(2014)\citenamefont {Feng},
  \citenamefont {Viebahn},\ and\ \citenamefont {Dijkstra}}]{GRL:GRL52004}%
  \BibitemOpen
  \bibfield  {author} {\bibinfo {author} {\bibfnamefont {Q.~Y.}\ \bibnamefont
  {Feng}}, \bibinfo {author} {\bibfnamefont {J.~P.}\ \bibnamefont {Viebahn}}, \
  and\ \bibinfo {author} {\bibfnamefont {H.~A.}\ \bibnamefont {Dijkstra}},\
  }\href@noop {} {\bibfield  {journal} {\bibinfo  {journal} {Geophysical
  Research Letters}\ }\textbf {\bibinfo {volume} {41}},\ \bibinfo {pages}
  {6009} (\bibinfo {year} {2014})}\BibitemShut {NoStop}%
\bibitem [{AVI(2013)}]{AVISO2013}%
  \BibitemOpen
  \href@noop {} {\bibfield  {journal} {\bibinfo  {journal} {CLS-DOS-NT-06-034}\
  }\textbf {\bibinfo {volume} {3rev 6}},\ \bibinfo {pages} {062806} (\bibinfo
  {year} {2013})}\BibitemShut {NoStop}%
\bibitem [{\citenamefont {Cimponeriu}\ \emph {et~al.}(2004)\citenamefont
  {Cimponeriu}, \citenamefont {Rosenblum},\ and\ \citenamefont
  {Pikovsky}}]{Cimponeriu2004}%
  \BibitemOpen
  \bibfield  {author} {\bibinfo {author} {\bibfnamefont {L.}~\bibnamefont
  {Cimponeriu}}, \bibinfo {author} {\bibfnamefont {M.}~\bibnamefont
  {Rosenblum}}, \ and\ \bibinfo {author} {\bibfnamefont {A.}~\bibnamefont
  {Pikovsky}},\ }\href@noop {} {\bibfield  {journal} {\bibinfo  {journal}
  {Phys. Rev. E}\ }\textbf {\bibinfo {volume} {70}},\ \bibinfo {pages} {046213}
  (\bibinfo {year} {2004})}\BibitemShut {NoStop}%
\bibitem [{\citenamefont {Toshiaki}\ \emph {et~al.}(2008)\citenamefont
  {Toshiaki}, \citenamefont {Roundy},\ and\ \citenamefont
  {Kiladis}}]{Toshiaki2008}%
  \BibitemOpen
  \bibfield  {author} {\bibinfo {author} {\bibfnamefont {S.}~\bibnamefont
  {Toshiaki}}, \bibinfo {author} {\bibfnamefont {P.}~\bibnamefont {Roundy}}, \
  and\ \bibinfo {author} {\bibfnamefont {G.}~\bibnamefont {Kiladis}},\
  }\href@noop {} {\bibfield  {journal} {\bibinfo  {journal} {J. Phys.
  Oceanogr.}\ }\textbf {\bibinfo {volume} {38}},\ \bibinfo {pages} {921}
  (\bibinfo {year} {2008})}\BibitemShut {NoStop}%
\bibitem [{\citenamefont {Eisenman}\ \emph {et~al.}(2005)\citenamefont
  {Eisenman}, \citenamefont {Yu},\ and\ \citenamefont
  {Tziperman}}]{Eisenman2005}%
  \BibitemOpen
  \bibfield  {author} {\bibinfo {author} {\bibfnamefont {I.}~\bibnamefont
  {Eisenman}}, \bibinfo {author} {\bibfnamefont {L.}~\bibnamefont {Yu}}, \ and\
  \bibinfo {author} {\bibfnamefont {E.}~\bibnamefont {Tziperman}},\ }\href@noop
  {} {\bibfield  {journal} {\bibinfo  {journal} {Journal of Climate}\ }\textbf
  {\bibinfo {volume} {18}},\ \bibinfo {pages} {5224} (\bibinfo {year}
  {2005})}\BibitemShut {NoStop}%
\bibitem [{\citenamefont {Benestad}\ \emph {et~al.}(2002)\citenamefont
  {Benestad}, \citenamefont {Sutton},\ and\ \citenamefont
  {Anderson}}]{QJ:QJ200212858211}%
  \BibitemOpen
  \bibfield  {author} {\bibinfo {author} {\bibfnamefont {R.~E.}\ \bibnamefont
  {Benestad}}, \bibinfo {author} {\bibfnamefont {R.~T.}\ \bibnamefont
  {Sutton}}, \ and\ \bibinfo {author} {\bibfnamefont {D.~L.~T.}\ \bibnamefont
  {Anderson}},\ }\href@noop {} {\bibfield  {journal} {\bibinfo  {journal}
  {Quarterly Journal of the Royal Meteorological Society}\ }\textbf {\bibinfo
  {volume} {128}},\ \bibinfo {pages} {1277} (\bibinfo {year}
  {2002})}\BibitemShut {NoStop}%
\bibitem [{\citenamefont {Mosquera-Vsquez}\ \emph {et~al.}(2014)\citenamefont
  {Mosquera-Vsquez}, \citenamefont {Dewitte},\ and\ \citenamefont
  {Illig}}]{JGRC:JGRC20892}%
  \BibitemOpen
  \bibfield  {author} {\bibinfo {author} {\bibfnamefont {K.}~\bibnamefont
  {Mosquera-Vsquez}}, \bibinfo {author} {\bibfnamefont {B.}~\bibnamefont
  {Dewitte}}, \ and\ \bibinfo {author} {\bibfnamefont {S.}~\bibnamefont
  {Illig}},\ }\href@noop {} {\bibfield  {journal} {\bibinfo  {journal} {Journal
  of Geophysical Research: Oceans}\ }\textbf {\bibinfo {volume} {119}},\
  \bibinfo {pages} {6605} (\bibinfo {year} {2014})}\BibitemShut {NoStop}%
\bibitem [{\citenamefont {Kessler}(1990)}]{Kessler1990}%
  \BibitemOpen
  \bibfield  {author} {\bibinfo {author} {\bibfnamefont {W.~S.}\ \bibnamefont
  {Kessler}},\ }\href@noop {} {\bibfield  {journal} {\bibinfo  {journal}
  {Journal of Geophysical Research}\ }\textbf {\bibinfo {volume} {95}},\
  \bibinfo {pages} {5183} (\bibinfo {year} {1990})}\BibitemShut {NoStop}%
\bibitem [{\citenamefont {Chelton}\ and\ \citenamefont
  {Schlax}(1996)}]{Chelton1996}%
  \BibitemOpen
  \bibfield  {author} {\bibinfo {author} {\bibfnamefont {D.~B.}\ \bibnamefont
  {Chelton}}\ and\ \bibinfo {author} {\bibfnamefont {M.~G.}\ \bibnamefont
  {Schlax}},\ }\href@noop {} {\bibfield  {journal} {\bibinfo  {journal}
  {Science}\ }\textbf {\bibinfo {volume} {272}},\ \bibinfo {pages} {234}
  (\bibinfo {year} {1996})}\BibitemShut {NoStop}%
\bibitem [{\citenamefont {Willett}\ \emph {et~al.}(2006)\citenamefont
  {Willett}, \citenamefont {Leben},\ and\ \citenamefont
  {Lav\'{i}n}}]{Willett2006}%
  \BibitemOpen
  \bibfield  {author} {\bibinfo {author} {\bibfnamefont {C.~S.}\ \bibnamefont
  {Willett}}, \bibinfo {author} {\bibfnamefont {R.~R.}\ \bibnamefont {Leben}},
  \ and\ \bibinfo {author} {\bibfnamefont {M.~F.}\ \bibnamefont {Lav\'{i}n}},\
  }\href@noop {} {\bibfield  {journal} {\bibinfo  {journal} {Progress in
  Oceanography}\ }\textbf {\bibinfo {volume} {69}},\ \bibinfo {pages} {218}
  (\bibinfo {year} {2006})}\BibitemShut {NoStop}%
\bibitem [{\citenamefont {Lyman}\ \emph {et~al.}(2005)\citenamefont {Lyman},
  \citenamefont {Chelton}, \citenamefont {Deszoeke},\ and\ \citenamefont
  {Samelson}}]{Lyman2005}%
  \BibitemOpen
  \bibfield  {author} {\bibinfo {author} {\bibfnamefont {J.~M.}\ \bibnamefont
  {Lyman}}, \bibinfo {author} {\bibfnamefont {D.~B.}\ \bibnamefont {Chelton}},
  \bibinfo {author} {\bibfnamefont {R.~A.}\ \bibnamefont {Deszoeke}}, \ and\
  \bibinfo {author} {\bibfnamefont {R.~M.}\ \bibnamefont {Samelson}},\
  }\href@noop {} {\bibfield  {journal} {\bibinfo  {journal} {J. Phys.
  Oceanogr.}\ }\textbf {\bibinfo {volume} {35}},\ \bibinfo {pages} {232}
  (\bibinfo {year} {2005})}\BibitemShut {NoStop}%
\bibitem [{\citenamefont {Ludescher}\ \emph {et~al.}(2014)\citenamefont
  {Ludescher}, \citenamefont {Gozolchiani}, \citenamefont {Bogachev},
  \citenamefont {Bunde}, \citenamefont {Havlin},\ and\ \citenamefont
  {Schellnhuber}}]{Ludescher11022014}%
  \BibitemOpen
  \bibfield  {author} {\bibinfo {author} {\bibfnamefont {J.}~\bibnamefont
  {Ludescher}}, \bibinfo {author} {\bibfnamefont {A.}~\bibnamefont
  {Gozolchiani}}, \bibinfo {author} {\bibfnamefont {M.~I.}\ \bibnamefont
  {Bogachev}}, \bibinfo {author} {\bibfnamefont {A.}~\bibnamefont {Bunde}},
  \bibinfo {author} {\bibfnamefont {S.}~\bibnamefont {Havlin}}, \ and\ \bibinfo
  {author} {\bibfnamefont {H.~J.}\ \bibnamefont {Schellnhuber}},\ }\href@noop
  {} {\bibfield  {journal} {\bibinfo  {journal} {Proceedings of the National
  Academy of Sciences}\ }\textbf {\bibinfo {volume} {111}},\ \bibinfo {pages}
  {2064} (\bibinfo {year} {2014})}\BibitemShut {NoStop}%
\end{thebibliography}
%

\end{document}